\documentclass[twocolumn,english,aps,pra,superscriptaddress,showpacs,tightenlines]{revtex4}
\usepackage[T1]{fontenc}
\usepackage[latin9]{inputenc}
\usepackage{amsmath}
\usepackage{amssymb}
\usepackage{graphicx}
\usepackage{wasysym}
\usepackage{esint}
\usepackage{wasysym}
\usepackage{babel}
\begin{document}

\title{Spontaneous Decoherence of Coupled Harmonic Oscillators Confined
in A Ring}

\author{Z. R. Gong}

\affiliation{College of Physics and Energy, Shenzhen University, Shenzhen 518060,
P. R. China}

\affiliation{Beijing Computational Science Research Center, Beijing 100084, China}

\author{Z. W. Zhang}

\affiliation{College of Physics and Energy, Shenzhen University, Shenzhen 518060,
P. R. China}

\author{D. Z. Xu}

\affiliation{School of Physics, Beijing Institute of Technology, Beijing 100081,
China}

\affiliation{Beijing Computational Science Research Center, Beijing 100084, China}

\author{N. Zhao}

\affiliation{Beijing Computational Science Research Center, Beijing 100084, China}

\author{C. P. Sun}

\affiliation{Beijing Computational Science Research Center, Beijing 100084, China}

\date{\today}

\begin{abstract}
We study the spontaneous decoherence of the coupled harmonic oscillators
confined in a ring container, where the nearest-neighbor harmonic
potentials are taken into consideration. Without any external symmetry
breaking field or surrounding environment, the quantum superposition
state prepared in the relative degrees of freedom gradually loses
its quantum coherence spontaneously. This spontaneous decoherence
is interpreted by the hidden couplings between the center-of-mass
and relative degrees of freedoms, which actually originates from the
symmetries of the ring geometry and corresponding nontrivial boundary
conditions. Especially, such spontaneous decoherence completely vanishes
at the thermodynamical limit because the nontrivial boundary conditions
become trivial Born-von Karman boundary conditions when the perimeter
of the ring container tends to infinity. Our investigation shows that
a thermal macroscopic object with certain symmetries has chance to
degrade its quantum properties even without applying an external symmetry
breaking field or surrounding environment.
\end{abstract}


\maketitle

\section{\label{sec:one}INTRODUCTION}

Quantum decoherence has been a subject of active research since the
quantum mechanics was established~\cite{Zurek91}. The revival of
the studies of the decoherence as a hot subject merits from the development
of the science and technology of the quantum information. As the physical
states in quantum mechanics are described by the superposition of
some eigenstates, the coherence existing between different eigenstates
is the important criteria for that whether the quantum properties
of the system remain or not. In this sense, the quantum decoherence
explains the emergence of the classical limit of a system with quantum
nature, which apparently determines the quantum-classical boundary~\cite{Haroche98,Zurek03,Bennett00,Quan06a}.

In the first place, quantum decoherence was named for the collapse
of the wave function in the Copenhagan interpretation~\cite{Howard04}.
In stead of generating actual wave function collapse, it only gives
the appearance of the wave function collapse. Nowadays, the studies
of the decoherence focus on the quantum correlation between the system
and its environment~\cite{Joos85,Zeh70,Zurek81,Zhou02}. As commonly
understood, the decoherence process can be viewed as that the quantum
system loses information into its environment. Mathematically, losing
information in decoherence process can be defined by the disappearance
of the off-diagonal elements of the system's reduced density matrix.
A perfect decoherence process requires that the environment approaches
its thermodynamic limit, whose infinite degrees of freedom guarantee
the infinitely long recurrence time of the decoherence process~\cite{Sun93,Zurek94,Sun01,Zeh01,Joos03,Schlosshauer05,Xue06}.

To reveal the mechanism of the quantum decoherence, Heisenberg introduced
a random phase factor according to the uncertainty principle. This
phase factor also results in the randomness of the coefficients of
the off-diagonal elements of the system's reduced density matrix,
whose average on time tends to zero eventually. However, the uncertainty
principle is not the only mechanism to cause decoherence, which has
been verified experimentally~\cite{Durr98,Arndt99}. Generally speaking,
the random factor originally comes from the interaction between the
quantum system and its environment. In contrast of the external environment
mentioned above, we are more interested in an internal one~\cite{Ommes94,Zhang02}.
For the most quantum systems, only some subspaces of the complete
Hilbert space of the system are concentrated on, whose adjoint space
can be regarded as the ``internal'' environment with interaction
between these two spaces such as the spin-orbit interaction, the electron-phonon
interaction and so on. Instead of infinite degrees of freedom the
external environment has, the internal environment only possesses
a few degrees of freedom.

Previous theoretical research indicated that due to the spontaneous
symmetry breaking~\cite{Wezel05,Wezel06,Wezel08a,Wezel08b} in association
with quantum phase transition~\cite{Quan06b}, the quantum decoherence
emerges in the multi-particle system when a small but finite symmetry
breaking field was added to a closed symmetric quantum system. Such
decoherence is called ``intrinsic decoherence'' because there is
no usual environment at all. When the symmetry is broken, a serious
of thin spectrum emerge in the vicinity of the original energy levels.
The subtle energy differences of the thin spectrum actually results
in the spontaneous decoherence. Recently, researchers show than the
spontaneous decoherence also can be induced by gravitational time
dilation~\cite{Pikovski15a,Pikovski15b,Gooding15}.

In this paper, we shall study the spontaneous decoherence of closed
multi-particle system without symmetry breaking. Considering $N$
coupled harmonic oscillators confined in a ring container, the Hamiltonian
can be decoupled into one center-of-mass motion and $N-1$ relative
motions. It is essential that the harmonic potentials between oscillators
are periodically repeated because of the ring configuration. Such
bosonic multi-particle system possesses $U(1)\otimes C_{\mathrm{N}}$
symmetry, where the continuous $U(1)$ symmetry and discrete $C_{\mathrm{N}}$
symmetry respectively relate the center-of-mass and relative motions'
symmetries. Then nontrivial boundary conditions emerge in order to
guarantee the single-valuedness of the wave function, which eventually
results in that the total energy spectrum not only depends on the
excitations of the relative motion, but also on the total momentum
corresponding to the center-of-mass motion. Similar to Aharonov-Bohm
effect, the nontrivial boundary conditions actually are equivalent
to applying an induced gauge fields~\cite{Yang61}. This hidden coupling
between the center-of-mass motion and relative motions introduces
a series of thin spectrum of the total momentum, which contributes
to the decoherence process of relative motions. If the center-of-mass
motion is not condensed to the state with single momentum, the spontaneous
decoherence process occurs in the superposition states in the relative
motions. Since there is no environment or symmetry breaking field
at all, the decoherence in our model is definitely intrinsic and its
dynamical process is spontaneous. The paradox of such spontaneous
decoherence is the information represented by the quantum coherence
is mysteriously missing in a completely closed system. The key point
to explain this is that the center-of-mass motion actually
acting like a surrounding environment to the relative motions we concentrate
on. The information is only transferred from the subspace of the complete
Hilbert space into its adjoint space.

This article is arranged as follows. We describe the multi-particle
model and derive the nontrivial boundary conditions in Sec. II. Then
the explicit total energy spectrum including all the thin spectrum
is obtained in Sec. III. In Sec. IV, we demonstrate how the thin spectrum
contributes to the dynamic decoherence process. We conclude in Sec
V.

\section{COUPLED HARMONIC OSCILLATORS CONFINED IN A RING CONTAINER}

\subsection{Model setup}

\begin{figure}[ptb]
\begin{centering}
\includegraphics[bb=29 79 553 801,clip,width=3.5in]{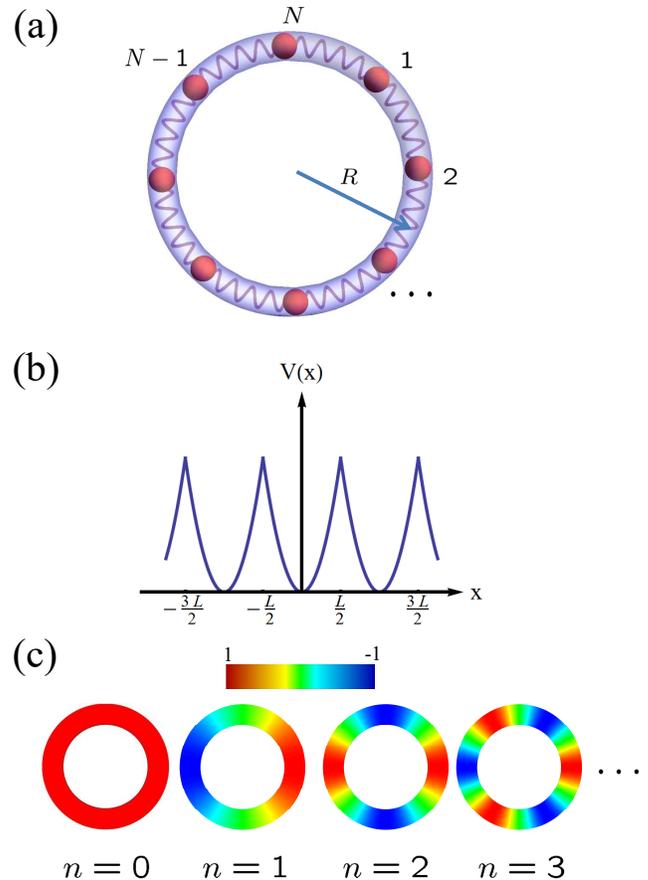}
\par\end{centering}

\caption{(Color online) (a) Schematic illustration of the coupled harmonic
oscillators confined in a ring container. Here, the light blue torus
and the red spheres represent the ring container with the perimeter
$R$ and the oscillators, respectively. (b) Schematic illustration
of the periodic harmonic potential. Here, $x$ denotes any displacement
difference between nearest neighbor oscillators. (c) Schematic illustration
of the real part of the center-of-mass motion wavefunction versus
the displacement $X_{0}$ for the first four quantum number $n$.}

\label{fig:fig1}
\end{figure}


To investigate the mechanism of the decoherence due to symmetries
of system, we consider a bosonic multi-particle system confined in
a ring container (Fig. 1(a)), which is modeled as $N$ coupled harmonic
oscillators with Hamiltonian
\begin{equation}
\hat{H}=\sum_{j=1}^{N}\left[\frac{\hat{p}_{j}^{2}}{2m}+V(\hat{x}_{j}-\hat{x}_{j+1})\right],\label{eq:2-1}
\end{equation}
where
\begin{equation}
V(\hat{x}_{j}-\hat{x}_{j+1})=\frac{\kappa}{2}\left(\hat{x}_{j}-\hat{x}_{j+1}\right)^{2}\label{eq:2-2}
\end{equation}
are harmonic potentials between nearest neighbor oscillators. Here,
$\hat{p}_{j}$ and $\hat{x}_{j}$ are the momentum and the displacement
of the $j-$th oscillator. For the sake of simplicity, the oscillator
mass $m$ and the spring constant $\kappa$ are supposed to be identical
for all oscillators, and the system is considered as one-dimensional
since the cross section radius of the ring container is much smaller
than the radius of the ring $R$. If all the oscillators only vibrate
in the vicinity of their equilibrium positions, it is the textbook
example of the phonons in the solid state physics with Born-von Karman
boundary condition. However, in the present situation, the oscillators
can potentially amove far away from their equilibrium positions if
their kinetic energies are sufficiently large. In this case, the harmonic
potentials becomes periodic as
\begin{equation}
V(x_{j}-x_{j+1}+nL)=V(x_{j}-x_{j+1})\label{eq:2-3}
\end{equation}
when any of the displacement difference between the nearest neighbor
oscillators is augmented by $nL$ ($n$ is an integer). This periodic
potential is schematically plotted in Fig. 1(b). Here, $L=2\pi R$
is the perimeter of the ring container. Since the harmonic potentials
only involve the displacement difference between the nearest neighbor
oscillators, the above coupled-oscillator system can be decoupled
to $N$ oscillators, which correspond to one center-of-mass motion
and $(N-1)$ relative motions.

To decouple the system into $N$ oscillators, we successively perform
the Fourier transformation as $\left(\hat{q}_{j}=\hat{p}_{j},\hat{x}_{j}\right)$
\begin{equation}
\hat{Q}_{k}=\begin{cases}
\sqrt{\frac{2}{N}}\sum_{j=1}^{N}\hat{q}_{j}\cos\left(\frac{2\pi kj}{N}\right), & 1\leq k\leq\frac{N}{2},\\
\sqrt{\frac{2}{N}}\sum_{j=1}^{N}\hat{q}_{j}\sin\left(\frac{2\pi kj}{N}\right), & k>\frac{N}{2},
\end{cases}\label{eq:2-4}
\end{equation}
where $\hat{Q}_{k}=\hat{P}_{k},\hat{X}_{k}$ $(k=1,\cdots,N-1)$ are
the momentums and the displacements of the $N-1$ independent relative
motions. Besides the relative motions, there is unique center-of-mass
motion, whose momentum and displacement are described as $\hat{P}_{0}=\sum_{j=1}^{N}\hat{p}_{j}$
and $\hat{X}_{0}=1/N\sum_{j=1}^{N}\hat{x}_{j}.$ We introduce the
different forms for momentums and the displacements when $1\leq k\leq N/2$
and $k>N/2$ in order to guarantee that they are still Hermitian operators
and satisfy the standard commutaion relation $\left[\hat{X}_{k},\hat{P}_{k'}\right]=i\hbar\delta_{k,k'}$.
After the Fourier transformation, the Hamiltonian becomes $N$ decoupled
harmonic oscillators as \begin{subequations}
\begin{eqnarray}
\hat{H} & = & \hat{H}_{0}+\sum_{k=1}^{N-1}\hat{H}_{k}\text{,}\label{eq:2-5-1}\\
\hat{H}_{k} & = & \frac{\hat{P}_{k}^{2}}{2m}+\frac{\kappa}{2}\left(2\sin\frac{\pi k}{N}\right)^{2}\hat{X}_{k}^{2}.\label{eq:2-5-2}
\end{eqnarray}
\end{subequations} It should be indicated that the zero-th Hamiltonian
\begin{equation}
\hat{H_{0}}=\frac{\hat{P}_{0}^{2}}{4mN}=\frac{1}{2mN}\left(\sum_{j=1}^{N}\hat{p}_{j}\right)^{2}\label{eq:2-6}
\end{equation}
describes the center-of-mass motion of the multiple particle system,
which is regarded as a whole carrying a kinetic energy associated
with total mass of the system. The rest part of the Hamiltonian $\hat{H}_{R}=\hat{H}-\hat{H}_{0}=\sum_{k=1}^{N-1}\hat{H}_{k}$
describes the decoupled $N-1$ relative motions. Obviously, each relative
mode is described by a periodic harmonic oscillator. Although the
periodicities of these relative motions are no longer simply demonstrated,
the sum of all relative harmonic oscillators potentials still possesses
the periodicities shown in Eq. (\ref{eq:2-3}). By solving the eigenvalue
problem of the system, we can obtain the thin spectrum which plays
essential role in our spontaneous quantum decoherence process.

\subsection{Origin of the thin spectrum}

Although the center-of-mass motion and relative motions seems independent
with each other in the Hamiltonian, there is a hidden coupling between
them due to the symmetry of the system. For a given quantum system,
the energy spectrum and eigen-wavefunctions are not only governed
by its Hamiltonian, but also determined by the boundary conditions
which depend on the symmetries of the system~\cite{Yang61}. We will
find the boundary conditions for our system as follows.

We first analyse the existing symmetries of the system shown in Fig.1(a).
If all the oscillator displacements $x_{j}(j=1,\ldots,N)$ are augmented
by the same increment $\delta x$, Hamiltonian keeps unchanged, which
means the system possesses $U(1)$ symmetry. Since the Hamiltonian
has been decoupled as Eq. (\ref{eq:2-5-1}), the eigenstate of the
system is obtained as
\begin{equation}
\Psi\left(\mathbf{X}\right)=\exp\left(\frac{i}{\hbar}P_{0}X_{0}\right)\chi\left(\mathbf{X}\right),\label{eq:2-7}
\end{equation}
where the plane wave $\exp\left(iP_{0}X_{0}/\hbar\right)$ and the
product state
\begin{equation}
\chi\left(\mathbf{\mathbf{X}}\right)=\prod\limits _{j=1}^{N-1}\chi_{j}\left(X_{j}\right)\label{eq:2-8}
\end{equation}
describes the center-of-mass motion and the relative motions, respectively.
Here, the vector $\mathbf{X}=\{X_{1},X_{2},\ldots,X_{N-1}\}$ represents
the displacements of relative motions as well as $\mathbf{x}=\{x_{1},x_{2},\ldots,x_{N-1}\}$
is the displacements of original oscillators. They are linked by the
a linear transformation as $\mathbf{X}=M\mathbf{x}$, where the transformation
matrix is determined by Eq. (\ref{eq:2-4}). If all the oscillator
displacements $x_{j}(j=1,\ldots,N)$ are augmented by the same increment
$\mu L$ ($\mu$ is integer), the relative motions keep unchanged
because all the relative displacements are unchanged, but there is
an additional phase to the center-of-mass motion wave function
\begin{equation}
\Psi'\left(\mathbf{\mathbf{X}}\right)=\exp\left(\frac{i}{\hbar}P_{0}\left(X_{0}+\mu L\right)\right)\chi\left(\mathbf{X}\right).\label{eq:2-9}
\end{equation}
The single-valuedness condition of the quantum mechanics requires
$\Psi'\left(\mathbf{\mathbf{X}}\right)=\Psi\left(\mathbf{X}\right),$
which leads to the quantized total momentum as ($n$ is integer)
\begin{equation}
P_{0}\left(n\right)=n\frac{\hbar}{R}.\label{eq:2-10}
\end{equation}
The real part of the plane waves of the center-of-mass motion versus
the displacement $X_{0}$ is depicted in Fig. 1(c). With the quantum
number $n$ increases, the nodes number of the real part of the center-of-mass
motion wavefunction also increases.

Besides this continuous symmetry, there is discrete symmetry due to
the periodicity of the harmonic potential shown in Eq. (\ref{eq:2-3}).
When any one of the displacement $x_{j}$ is augmented by $\mu L$,
the Hamiltonian is still unchanged. In this sense, the operation not
only introduces a similar phase to the center-of-mass motion as
\begin{equation}
\Psi'\left(\mathbf{\mathbf{\mathbf{X}}'}\right)=\exp\left[\frac{i}{\hbar}P_{0}\left(n\right)\left(X_{0}+\frac{1}{N}\mu L\right)\right]\chi\left(\mathbf{\mathbf{X}}'\right),\label{eq:2-11}
\end{equation}
but also change the displacements of the relative motions to $\mathbf{\mathbf{X}}'=\mathbf{X}+\mu L\boldsymbol{M}_{j_{0}}.$
Here, $\boldsymbol{M}_{j_{0}}$ is the column vector of the transformation
matrix $\mathbf{M}$. If we only focus on the additional phase of
the center-of-mass motion and substitute the quantized total momentum
in Eq. (\ref{eq:2-10}), the phase $\exp\left(i2\pi n\mu/N\right)$
actually only have $N$ possible values for $\mod[n\mu,N]=0,1,\ldots,N-1,$
where $\mod[x,y]$ gives the remainder on division of x by y. These
$N$ operations actually constitutes the $N$ elements of the $C_{N}$
group. Therefore, the total system symmetry group is $U(1)\otimes C_{\mathrm{N}}$.

To obtain the energy spectrum, the corresponding Schrodinger equation
is taken into consideration as
\begin{equation}
\hat{H}\Psi\left(\mathbf{X}\right)=E\left(n\right)\Psi\left(\mathbf{X}\right),\label{eq:2-12}
\end{equation}
where the eigen-energy contains the kinetic energy of center-of-mass
motion and the energies of the relative motions as
\begin{equation}
E\left(n,\alpha\right)=\frac{n^{2}\hbar^{2}}{2mNR^{2}}+\epsilon\left(\alpha\right).\label{eq:2-13}
\end{equation}
Here, we already have substituted the quantized total momentum into
the kinetic energy $P_{0}^{2}/2mN$. Since the total momentum commutes
with all displacements of relative motions as $\left[\hat{X}_{k},\hat{P}_{0}\right]=i\hbar\delta_{k,0}$,
the eigenstates describing the relative motions also satisfy the following
Schrodinger equation as
\begin{equation}
\hat{H}\chi\left(\mathbf{\mathbf{X}}\right)=\epsilon\chi\left(\mathbf{X}\right).\label{eq:2-14}
\end{equation}

Usually, the energy spectrum of the relative modes $\epsilon$ is
independent of the total momentum $P_{0}$ and the coherence of the
relative motion states can be maintained all the time. However, single-valuedness
condition requires the wavefunction in Eq. (\ref{eq:2-11}) is the
same as the wavefunction in Eq. (\ref{eq:2-7}), which leads to
\begin{equation}
\chi\left(\mathbf{X}\right)=\exp\left(i\mu\theta_{n}\right)\chi\left(\mathbf{X}+\mu L\boldsymbol{M}_{k_{0}}\right)\label{eq:2-15}
\end{equation}
with $\theta_{n}=2\pi n/N$ for any $k_{0}=1,2,\ldots,N-1$. Here,
the boundary conditions in Eq. (\ref{eq:2-15}) actually can guarantee
the single-valuedness condition for any $\mu$ as
\begin{align}
\chi\left(\mathbf{\mathbf{X}}\right) & =\exp\left(i\theta_{n}\right)\chi\left(\mathbf{X}+L\boldsymbol{M}_{k_{0}}\right)\nonumber \\
 & =\exp\left(i2\theta_{n}\right)\chi\left(\mathbf{X}+2L\boldsymbol{M}_{k_{0}}\right)\nonumber \\
 & =\ldots=\exp\left(i\mu\theta_{n}\right)\chi\left(\mathbf{X}+\mu L\boldsymbol{M}_{k_{0}}\right).\label{eq:2-16}
\end{align}
Obviously, $\theta_{n}$ depends on the total momentum $P_{0}$, which
eventually results in that the energy of relative motions $\epsilon\left(n,\alpha\right)$
becomes dependent of the total momentum. For different quantum number
$n$ of the total momentum, the group of the energy levels form the
thin spectrum, which plays the essential role in the spontaneous decoherence.
The Hamiltonian in the first place possesses the $C_{N}$ symmetry
implying periodic $\theta_{n}$ as $\theta_{n}$=$\theta_{n+\mu N}$,
therefore the thin spectrum is also periodic as $\epsilon\left(n,\alpha\right)=\epsilon\left(n+\mu N,\alpha\right).$
Since the Hamiltonian has inversion symmetry when $\mathbf{x}\rightarrow-\mathbf{x},$
which imply that the thin spectrum is even function of $n$ as $\epsilon\left(n,\alpha\right)=\epsilon\left(-n,\alpha\right).$

We will solve the energy spectrum of the relative motions from its
eigen-equation in Eq. (\ref{eq:2-14}) together with the nontrivial
boundary conditions in Eq. (\ref{eq:2-15}) in order to obtain the
thin spectrum depending on the quantum number $n$ of the total momentum
in the next section.

\section{THE TOTAL ENERGY SPECTRUM}

Since the harmonic oscillator potential for relative motions are still
periodic, according to the Floquet theorem~\cite{Magnus04}, the
$k-$th relative motion can be rewritten as
\begin{eqnarray}
\chi_{k}\left(X_{k}\right) & = & e^{iq_{k}X_{k}}u_{k}\left(X_{k}\right)\label{eq:3-1}
\end{eqnarray}
with wave vector $q_{k}$ and the periodic part $u_{k}\left(X_{k}\right).$
According to Eq. (\ref{eq:2-8}), the total relative motions are described
by the product state as
\begin{equation}
\chi\left(\mathbf{\mathbf{X}}\right)=e^{\sum_{k=1}^{N-1}iq_{k}X_{k}}\prod\limits _{k=1}^{N-1}u_{k}\left(X_{k}\right).\label{eq:3-2}
\end{equation}
In order to satisfy the boundary conditions as Eq. (\ref{eq:2-15}),
we calculate the wavefunction of all the relative motions when the
$j_{0}-$th oscillator displacement is augmented by $L$ as
\begin{eqnarray}
 &  & \chi\left(\mathbf{X}+L\boldsymbol{M}_{j_{0}}\right)\nonumber \\
 & = & e^{\sum_{k=1}^{N-1}iq_{k}\left(X_{k}+LM_{k_{0}}^{k}\right)}\prod\limits _{k=1}^{N-1}u_{k}\left(X_{k}+LM_{k_{0}}^{k}\right)\nonumber \\
 & = & e^{\sum_{k=1}^{N-1}iq_{k}LM_{k_{0}}^{k}}e^{\sum_{k=1}^{N-1}iq_{k}X_{k}}\prod\limits _{k=1}^{N-1}u_{k}\left(X_{k}+LM_{k_{0}}^{k}\right)\nonumber \\
 & = & e^{\sum_{k=1}^{N-1}iq_{k}LM_{k_{0}}^{k}}\chi\left(\mathbf{X}\right),\label{eq:3-3}
\end{eqnarray}
where $M_{k_{0}}^{k}$ are the elements of the vector $\boldsymbol{M}_{k_{0}}=\left(M_{k_{0}}^{1},M_{k_{0}}^{2},\ldots,M_{k_{0}}^{N-1}\right)$
and in the last step we apply the periodicity of the wavefunctions
$\left\{ u_{k}\left(X_{k}\right)\right\} $ as
\begin{equation}
\prod\limits _{k=1}^{N-1}u_{k}\left(X_{k}+LM_{k_{0}}^{k}\right)=\prod\limits _{k=1}^{N-1}u_{k}\left(X_{k}\right).\label{eq:3-4}
\end{equation}
In contrat with the boundary conditions in Eq. (\ref{eq:2-15}), we
actually obtain the constrains for the wave vectors $\{q_{k}\}$ as
\begin{equation}
LM_{k_{0}}^{k}q_{k}+\theta_{n}=0,\label{eq:3-5}
\end{equation}
which should be satisfied for any $k_{0}.$ The $N-1$ constrains
completely determine the wave vectors $\{q_{k}\}$. In the vector
form, it can be rewritten as
\begin{equation}
L\mathbf{M}\mathbf{q}+\theta_{n}\mathbf{I}=0,\label{eq:3-6}
\end{equation}
where $\mathbf{q}=(q_{1},q_{2,}\ldots,q_{N-1})^{T}$ and $\mathbf{I}=(1,1,\ldots,1)^{T}$.
The solution is straightforwardly obtained as (see Appendix A)
\begin{equation}
q_{j}=\begin{cases}
q, & 1\leq k\leq\frac{N-1}{2},\\
0, & \frac{N+1}{2}\leq k\leq N-1,
\end{cases}\label{eq:3-7}
\end{equation}
for odd number $N$ and
\begin{equation}
q_{j}=\begin{cases}
q, & 1\leq k\leq\frac{N}{2}-1,\\
\frac{q}{2}, & k=\frac{N}{2},\\
0, & \frac{N}{2}+1\leq k\leq N-1,
\end{cases}\label{eq:3-8}
\end{equation}
for even number $N$ with $q=\frac{\sqrt{2}n}{\sqrt{N}R}.$ It indicates
that for those relative motions with $k<N/2$ the wave vectors $q$
are exactly same, which is proportional to the quantum number $n$
as well as the total momentum $P_{0}\left(n\right).$ While for those
relative motions with $k>N/2$ the wave vectors vanish. In this sense,
the phase factor $\theta_{n}$ resulting from the total momentum now
is divided into individual phase factors of those relative motions
with $k\leq N/2$. Actually, the consequence of the nontrivial boundary conditions is adding an
additional phase factor in Eq. (\ref{eq:2-15}), which actually is
equivalent to introducing a gauge field onto the relative motions
(see Appendix B).

Therefore it is feasible to deal with single relative motion in order
to obtain the corresponding energy spectrum once the individual periodicity
of the relative motion is determined. When the $j_{0}$-th oscillator's
displacement is augmented by $\mu L$, the change of the relative
motion displacements is $\mathbf{\mathbf{X}}'=\mathbf{X}+\mu L\boldsymbol{M}_{k_{0}}$
and the periodic part of the wavefunction $u_{k}\left(X_{k}\right)$
satisfies
\begin{eqnarray}
u_{k}\left(X_{k}\right) & = & u_{k}\left(X_{k}+LM_{k_{0}}^{k}\right).\label{eq:3-9}
\end{eqnarray}
Since we can permutate the indices of the original oscillators such
as $\{k_{0},k_{0}+1,\ldots,N,1,2,\ldots,k_{0}-1\}\rightarrow\{1,2,\ldots,N\}$
in order to always augment the first oscillator's displacement, the
periodicities of those relative motions are considered as $u_{k}\left(X_{k}\right)=u_{k}\left(X_{k}+LM_{1}^{k}\right).$
In this sense, we can solve the Schrodinger equation
\begin{equation}
\hat{H_{k}}\chi_{k}\left(X_{k}\right)=\epsilon_{k}\chi_{k}\left(X_{k}\right)\label{eq:3-10}
\end{equation}
and corresponding boundary conditions, which require both the wavefunction
and derivative of the wavefunction is continuous as\begin{subequations}
\label{eq:3-11}
\begin{eqnarray}
\chi_{k}\left(-\frac{L}{2}M_{1}^{k}\right) & = & e^{iq_{k}LM_{1}^{k}}\chi_{k}\left(\frac{L}{2}M_{1}^{k}\right),\label{eq:3-11-1}\\
\left.\frac{d}{dX_{k}}\chi_{k}\left(X_{k}\right)\right|_{X_{k}=-\frac{L}{2}M_{1}^{k}} & = & e^{iq_{k}LM_{1}^{k}}\left.\frac{d}{dX_{k}}\chi_{k}\left(X_{k}\right)\right|_{X_{k}=\frac{L}{2}M_{1}^{k}}.\label{eq:3-11-2}
\end{eqnarray}
\end{subequations}The energy spectrum depends on quantum number $n$
can be approximately obtained as (see Appendix C)

\begin{equation}
\epsilon_{k}\left(n,\alpha\right)=\left(\frac{1}{2}+\alpha+\delta_{k}\left(n,\alpha\right)\right)\hbar\omega_{k}\label{eq:3-12}
\end{equation}
with the frequency of the oscillator of the $k$-th relative motion
$\omega_{k}=4\sqrt{\kappa/m}\sin\left(\pi k/N\right)$. The explicit
form of the total-momentum dependent term $\delta_{k}\left(n,\alpha\right)$
can be found in Appendix C.

The total thin spectrum is the sum of all the energies of the relative
motions as $\epsilon\left(n,\alpha\right)=\sum_{k=1}^{N-1}\epsilon_{k}\left(n,\alpha\right).$
The schematics of the the spectrum is depicted in Fig.2, which is
almost quadratic of the $n$ and linear of $\alpha$. The subtle difference
between different thin spectra with different excitation quantum number
of the relative modes $\alpha$ usually still depends on the total
momentums, which leads to the decoherence of the relative modes. The
details of such decoherence process will be discussed in the next
section.

\begin{figure}[ptb]
\begin{centering}
\includegraphics[bb=34 300 576 739,clip,width=3.5in]{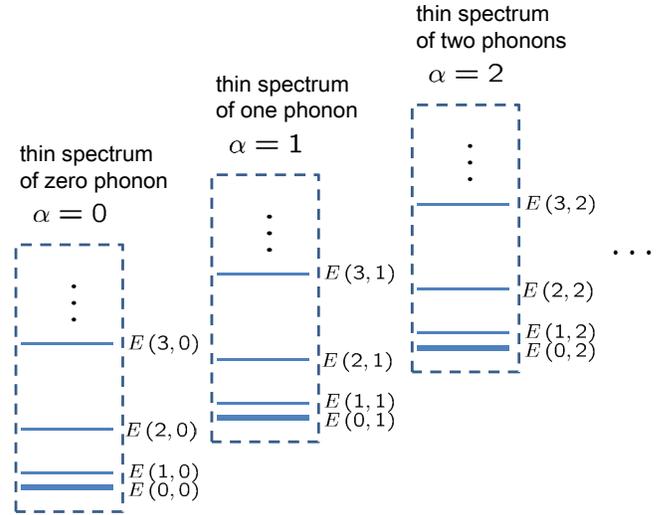}
\par\end{centering}

\caption{Schematic illustration of the total energy spectrum $E\left(n,\alpha\right)$.
The thin spectrum is almost quadratic of the quantum number $n$ of
the total momentum, and linear of the quantum number $\alpha$ of
the relative motions.}

\label{fig:fig2}
\end{figure}


\section{DECOHERENCE OF THE RELATIVE MOTIONS}

\subsection{Decoherence factor}

To explore the decoherence of the relative modes caused by the thin
spectrum, we consider the dynamics of an actual qubit of the multi-particle
system. The qubit is chosen as $(a\left|0\right\rangle +b\left|1\right\rangle )\otimes\left|n\right\rangle $with
the ground state of the relative modes $\left|\alpha=0\right\rangle $,
the first excitation state of the relative modes $\left|\alpha=1\right\rangle $
and the center-of-mass state $\left|n\right\rangle $ (see fig.2).
If the multi-particle system condensates on the BEC state with a single
momentum, which is equivalent to that $\left|n\right\rangle $ only
contains a single mode plane wave, the effect of the thin spectrum
is adding a phase factor to the off-diagonal elements of the reduced
density matrix of the relative modes and thus no decoherence process
occurs. However, in a relative high temperature such as $k_{B}T\gg\hbar^{2}/2mNR^{2},$
the center-of-mass state usually stays in thermal state as
\begin{equation}
\rho_{T}=\frac{1}{Z}\sum_{n=-\infty}^{\infty}e^{-\beta E\left(n,\alpha\right)}\left|n\right\rangle \left\langle n\right|\label{eq:4-1}
\end{equation}
for a macroscopic object with $\beta^{-1}=k_{B}T$, where the thin
spectrum is labeled by the quantum number $n$ of total momentum and
$\alpha$ of the relative motions as
\begin{eqnarray}
E\left(n,\alpha\right) & = & \frac{n^{2}\hbar^{2}}{2mNR^{2}}+\epsilon\left(n,\alpha\right)\label{eq:4-2}
\end{eqnarray}
and $Z=\sum_{n}e^{-\beta E\left(n,0\right)}$ is the partition function
corresponding to the product of the center-of-mass thermal state and
the ground state of the relative modes.

We prepare the initial state of the qubit on its ground state $\left|0\right\rangle $
and then apply a rotation to transform the ground state into $a\left|0\right\rangle +b\left|1\right\rangle .$
In this case, the initial density matrix is the product of the thermal
state density matrix and qubit one as
\begin{eqnarray}
\rho_{0} & = & \rho_{T}\otimes\rho_{Q}\nonumber \\
 & = & \frac{1}{Z}\sum_{n=-\infty}^{\infty}e^{-\beta E\left(n,0\right)}\left|n\right\rangle \left\langle n\right|\times\nonumber \\
 &  & (a\left|0\right\rangle +b\left|1\right\rangle )(a^{*}\left\langle 0\right|+b^{*}\left\langle 1\right|).\label{eq:4-3}
\end{eqnarray}

Since we have solved the total energy spectrum of the system, the
time evolution of the eigenstate $\left|n,\alpha\right\rangle \equiv\left|n\right\rangle \otimes\left|\alpha\right\rangle (\alpha=0,1)$
can be described by a time evolution operator as
\begin{equation}
U_{t}\left|n,\alpha\right\rangle =\exp\left[-\frac{i}{\hbar}E\left(n,\alpha\right)t\right]\left|n,\alpha\right\rangle .\label{eq:4-4}
\end{equation}
Then the time evolution of the density matrix is
\begin{eqnarray}
\rho_{t} & = & U_{t}\rho_{0}U_{t}^{\dagger}\nonumber \\
 & = & \frac{1}{Z}\sum_{n=-\infty}^{\infty}e^{-\beta E\left(n,0\right)}\left|n\right\rangle \left\langle n\right|\left(\left|a\right|^{2}\left|0\right\rangle \left\langle 0\right|+\left|b\right|^{2}\left|1\right\rangle \left\langle 1\right|\right.\nonumber \\
 &  & \left.+a^{*}be^{-\frac{i}{\hbar}(E\left(n,1\right)-E\left(n,0\right))t}\left|1\right\rangle \left\langle 0\right|+h.c.\right).\label{eq:4-5}
\end{eqnarray}
Tracing out the degree of freedom of the center-of-mass, we can define
the decoherence factor from the coefficients of the off-diagonal elements
as
\begin{equation}
F=\left|\frac{1}{Z}\sum_{n=-\infty}^{\infty}e^{-\beta E\left(n,0\right)}e^{-\frac{i}{\hbar}\Delta E\left(n\right)t}\right|\label{eq:4-6}
\end{equation}
with $\Delta E\left(n\right)=E\left(n,1\right)-E\left(n,0\right)$.
Obviously, the decoherence factor is equal or less then $1$, which
characterizes the completeness of the decoherence process. $F=1$
means the state has the same coherence as the initial quantum state,
$F<1$ means the decoherence occurs and the multi-particle system
becomes classical when $F=0.$

\subsection{Time scale of the decoherence at two limits}

Since the ground state is the product of the ground states of all
relative motions, namely $\left|0\right\rangle =\prod_{k=1}^{N-1}\otimes\left|0_{k}\right\rangle $,
the ground state energy
\begin{equation}
E\left(n,0\right)=\frac{n^{2}\hbar^{2}}{2mNR^{2}}+\sum_{k=1}^{N-1}\left(\frac{1}{2}+\delta_{k}^{n}\right)\hbar\omega_{k}\label{eq:4-7}
\end{equation}
is the summation of the ground state energy of all relative motions
and the kinetic energy of center-of-mass motion. Additionally, since
the first excited state is the state that $\left(N-1\right)$-th relative
motions remain at ground state and only the first relative motion
is excited to the excited state as $\left|1\right\rangle =\left|1_{1}\right\rangle \prod_{k=2}^{N-1}\otimes\left|0_{k}\right\rangle ,$
the energy difference in the decoherence factor actually only depends
on the energy level spacing of the ground state and the excited state
of the first relative motion, namely
\begin{align}
 & \Delta E\left(n\right)=\epsilon_{1}\left(n,1\right)-\epsilon_{1}\left(n,0\right)\nonumber \\
 & \approx-\hbar\omega_{1}\frac{g}{2}\cos\left(4\pi\frac{n}{N}\right),\label{eq:4-8}
\end{align}
where $g\hbar\omega_{1}=\Delta E\left(N/4\right)-\Delta E\left(0\right)$
is the maximum energy difference between thin spectrum. Here, we have
assumed the thin spectrum has the cosine type oscillating behavior
because it is periodic even function associating with the period $N/2$
of phase factor $\theta_{n}$. Under this approximation, the decoherence
factor in Eq. (\ref{eq:4-6}) can be written in a series of Bessel
functions as
\begin{align}
F & \approx\left|\frac{1}{Z}\int_{-\infty}^{\infty}e^{-\beta\Delta_{e}'n^{2}}e^{i\frac{g}{2}\omega_{1}t\cos\left(4\pi\frac{n}{N}\right)}dn\right|\nonumber \\
 & =\left|\frac{1}{Z}\int_{-\infty}^{\infty}e^{-\beta\Delta_{e}'n^{2}}\sum_{\gamma=-\infty}^{\infty}e^{i\alpha\left(4\pi\frac{n}{N}+\frac{\pi}{2}\right)}J_{\gamma}\left[\frac{g}{2}\omega_{1}t\right]dn\right|\nonumber \\
 & =\left|\sum_{\gamma=-\infty}^{\infty}J_{\gamma}\left(\frac{g}{2}\omega_{1}t\right)e^{i\gamma\frac{\pi}{2}}\exp\left(-\frac{4\pi^{2}\gamma^{2}}{N^{2}\beta\Delta_{e}'}\right)\right|.\label{eq:4-9}
\end{align}
Here, we have assumed the second term in $E\left(n,0\right)$ is quadratic
of $n$ as $\sum_{k=1}^{N-1}\delta_{k}^{n}\hbar\omega_{k}=\Delta_{e}n^{2}$
and $\Delta_{e}'=\Delta_{e}+\hbar^{2}/2mNR^{2}.$ We also have neglected
the $n$ independent term because they will vanish in the absolute
value of the Eq. (\ref{eq:4-6}).

Obviously for the first limit, if $4\pi^{2}/N^{2}\beta\Delta_{e}'\gg1$
the last term exponentially decays as $\gamma$ increases and eventually
only $\gamma=0$ term contributes to the decoherecne factor as $F=J_{0}\left(\frac{g}{2}\omega_{1}t\right).$
In this limit, the decoherence factor is independent of the temperature
and has an oscillating behavior associating with the $0$-th Bessel
function.

We can obtain the decoherence factor in another limit. Since the decoherence
factor in Eq.(\ref{eq:4-6}) basically is the integral of both the
Gaussian part and the dynamic phase, if the period of the dynamic
phase ($N/2$) is greater than the full width at half maximum (FWHM)
of the Gaussian part, only the first period of the thin spectrum contributes
to the decoherence factor. In this sense, the energy difference is
approximately linear one as

\begin{align}
\Delta E\left(n\right)= & \epsilon_{1}\left(n,1\right)-\epsilon_{1}\left(n,0\right)\nonumber \\
\approx & \frac{\Delta_{g}}{N}\left|n\right|\hbar\omega_{k}\label{eq:4-10}
\end{align}
with $\Delta g=g_{1}\left(1,1\right)-g_{1}\left(1,0\right)$. The
defininition of function $g_{1}\left(k,m\right)$ can be found in
Appendix C. The decoherence factor actually possesses an exponentially
decay behavior as
\begin{align}
F & \approx\left|\frac{1}{Z}\int_{-\infty}^{\infty}e^{-\beta\Delta_{e}'n^{2}}e^{-i\sqrt{\frac{\kappa}{m}}2\pi\frac{\Delta_{g}}{N^{2}}\left|n\right|t}dn\right|\nonumber \\
 & =e^{-\left(\frac{t}{\tau}\right)^{2}}\sqrt{1+\mathrm{Erfi}\left(\frac{t}{\text{\ensuremath{\tau}}}\right)^{2}},\label{eq:4-11}
\end{align}
where

\begin{equation}
\tau=\sqrt{\frac{\beta N^{4}m\Delta_{e}'}{\pi^{2}\Delta_{g}^{2}\kappa}},\label{eq:4-12}
\end{equation}
$\mathrm{Erfi}(t/\tau)$ is the imaginary error function and the summation
becomes a integral at high temperature such as $k_{B}T\gg\hbar^{2}/2mNR^{2}$.
The typical time scale of the decoherence is
\begin{equation}
\tau_{spon}=\sqrt{\frac{2\left(\pi-2\right)}{\pi}}\tau\approx0.85\tau.\label{eq:4-13}
\end{equation}
Since usually the $\Delta_{e}$ and $\Delta_{g}$ usually depend on
all other parameters such as $N$,$T$, $\kappa,$ $m$ and $R$ (see
Appendix C), it is hard to determine the exact dependence of the decoherence
factor on those parameters, and we will present numerical analysis
in the next subsection. Especially, if the lattice constant $R/N$
is unchanged while increasing the ring container radius $R$, the
decoherence tends to infinity, $\Delta_{e}$ and $\Delta_{g}$ both
tend to constant and thus the $\tau_{spon}$ is proportional to the
$\sqrt{R}$. This implies no spontaneous decoherence occurs in the
thermodynamical limit. This is consistent with the textbook example
of phonon in the solid state physics.

\subsection{Numerical results}
\begin{figure}[ptb]
\begin{centering}
\includegraphics[bb=17 450 572 785,clip,width=3.5in]{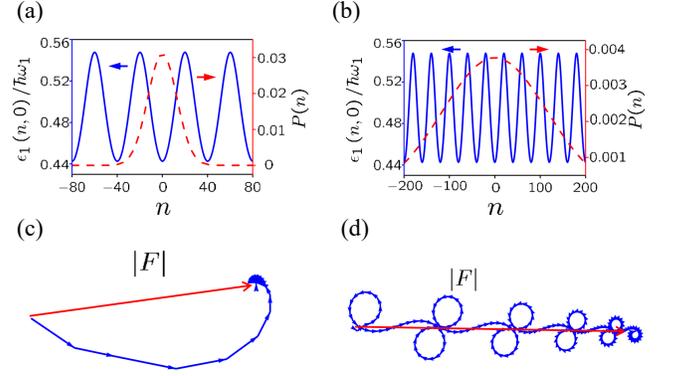}
\par\end{centering}

\caption{(a) (b) The typical thin spectrum $\epsilon_{1}\left(n,0\right)$
for the zero phonon of the first relative motion and the normalized
Gaussian part $P\left(n\right)$ versus the quantum number $n$. The
blue solid line and the red dashed line represent $\epsilon_{1}\left(n,0\right)$
and $P\left(n\right)$, respectively. The parameters are chosen as
$N=80,R=0.5\mu m,\kappa=10^{-13}N/s,m=40m_{p}$ with $m_{p}$ is mass
of proton. The temperature is $T=0.1\mu K$ for (a) and $T=8\mu K$
for (b). (c) and (d) are the typical vector summation pictures of
the decoherence factor $\left|F\right|$ respectively corresponds
to the cases (a) and (b). The red arrow and the blue arrows respectively
represent the decoherence factor $\left|F\right|$ and the successive
terms in decoherence factor. Obviously, for case (a) there is only
the single period of the thin spectrum contributing to the decoherence
factor. While for the case (b) there are multi-periods of the thin
spectrum contributing to the decoherence factor.}

\label{fig:fig3}
\end{figure}
The numerical calculations based on Eq. (~\ref{eq:4-6}) are present
in this section. The typical thin spectrum $\epsilon_{1}\left(n,0\right)$
for the zero phonon of the first relative motion and the normalized
Gaussian part
\begin{equation}
P\left(n\right)=\frac{1}{Z}e^{-\beta\left(\frac{n^{2}\hbar^{2}}{2mNR^{2}}\right)}\label{eq:4-14}
\end{equation}
in the decoherence factor versus the quantum number $n$ are depicted
in Fig. 3(a) and (b). The parameters are chosen as $N=80,R=0.5\mu m,\kappa=10^{-13}N/s,m=40m_{p}$
with $m_{p}$ the mass of proton. The temperature is $T=0.1\mu K$
for (a) and $T=8\mu K$ for (b). This mechanism is depicted in the
Fig. 3(c) and (d), where the each complex successive term in the summation
of the decoherence factor is regarded as a vector. In this sense of
the vector summation picture, the decoherence factor is the length
of the vector summation. There are three typical decoherence processes.
If all the phases of the vectors are the same, the coherence can be
maintained well. If the $N/4$ is larger than the full width at half
maximum of the Gaussian part, only the first period of the thin spectrum
contributes to the decoherence factor shown in Fig. 3(a) and (c).
While $N/4$ is smaller than the FWHM of the Gaussian part, the next
several periods of the thin spectrum also contributes to the decoherence
factor and usually it will elongate the decoherence time shown in
Fig. 3(b) and (d). Usually, the FWHM of the Gaussian part
\begin{equation}
n_{\mathrm{FWHM}}=\sqrt{\frac{2mNR^{2}}{\beta\hbar^{2}}}\label{eq:4-15}
\end{equation}
decreases when decreasing the temperature $T$, the particle mass
$m$, the particle number $N$ and the radius of the ring container
$R$. In this sense, we can define one parameter
\begin{equation}
r=\frac{n_{\mathrm{FWHM}}}{N/4}=4\sqrt{\frac{2mR^{2}}{\beta N\hbar^{2}}}\label{eq:4-16}
\end{equation}
to distinguish these two cases, where $r<1$ and $r>1$ respectively
corresponds to single and multi period contributions shown in Fig.
3(a) and Fig. 3 (b).

\begin{figure}[ptb]
\begin{centering}
\includegraphics[bb=16 119 452 775,clip,width=3in]{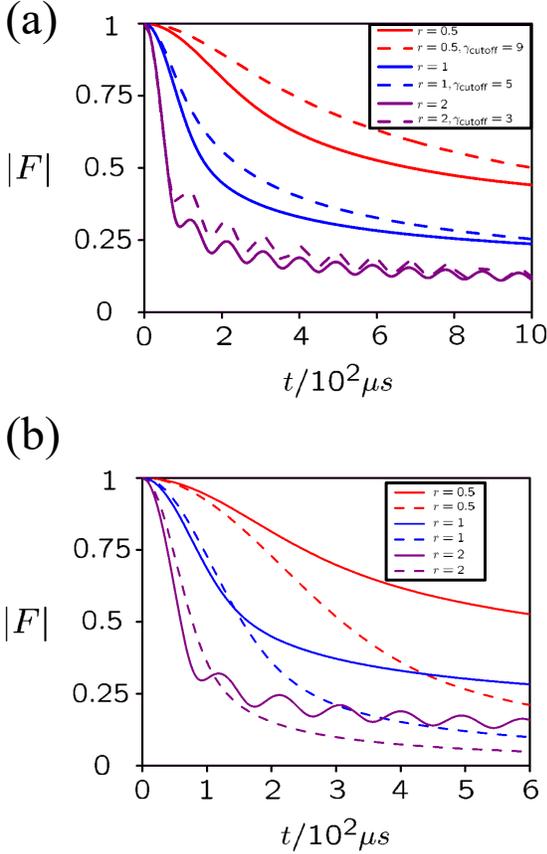}
\par\end{centering}

\caption{The decoherence factor obtained by the exact solution (solid lines)
and the approximate solution (dashed lines) based on (a) Eq.(\ref{eq:4-9}) and (b) Eq.(\ref{eq:4-11}). The parameters are chosen as $N=80,\kappa=10^{-13}N/s,R=0.5\mu m,m=40m_{p}$. The temperatures are respectively $T=483nK, 121nK,31nK$ to guarantee $r=0.5,1,2$ for red, blue and purple lines. And the cutoff $\gamma_\mathrm{cutoff}=9,5,3$ for red, blue and purple dashed lines respectively.}

\label{fig:fig4}
\end{figure}
\begin{figure}[ptb]
\begin{centering}
\includegraphics[bb=11 99 563 768,clip,width=3.5in]{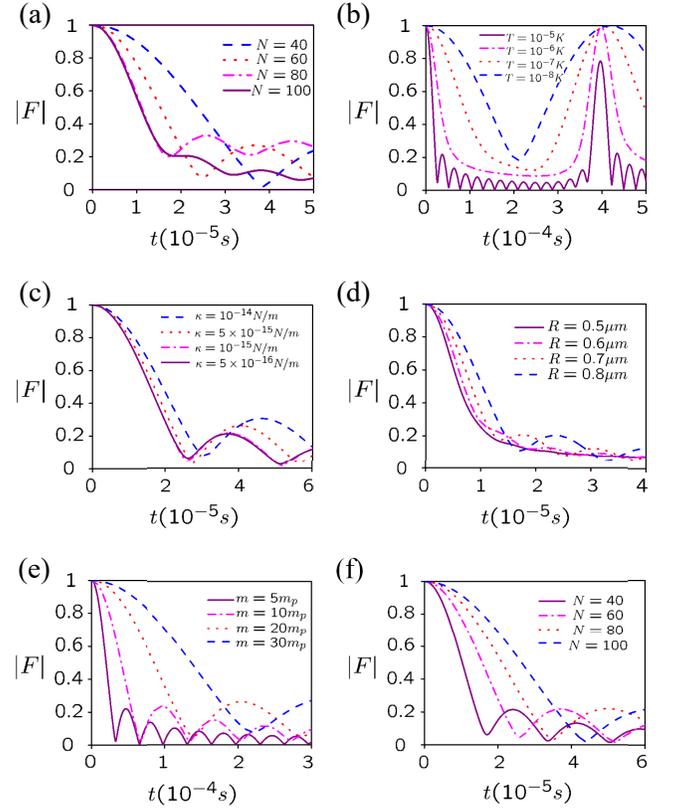}
\par\end{centering}

\caption{The evolutions of the decoherence factor for different (a)particle
number $N$, (b)the temperature $T$, (c)spring constant $\kappa$,
(d)the radius of the ring container $R$ and (e)the particle mass
$m$. For (f), the particle number $N$ and the ring container $R$
increase simultaneously in order to keep the linear mass density $\eta=N/2\pi R$
unchanged. The basic parameters are chosen as $N=80,T=10^{-5}K,\kappa=10^{-13}N/s,R=1\mu m,m=4m_{p}$
with $m_{p}$ is mass of proton. Basically, the decoherence time is
elongated for smaller particle number, lower temperature, stronger
spring constant, larger ring container, and heavier particle. If the
linear mass density $\eta=N/2\pi R\approx9.55\times10^{6}m^{-1}$
is kept unchanged and increase the particle number, the decoherenc
time is elongated instead of being shortened when only the particle
number is increased shown in (a). It implies that the spontaneous
decoherence vanishes at the thermodynamical limit.}

\label{fig:fig5}
\end{figure}

Eq. (\ref{eq:4-9}) is valid to describe the decoherence process
when the thin spectrum approximately has the cosine type oscillating
behavior. The contrast of the exactly decoherence factor obtained
from Eq. (\ref{eq:4-6}) and the approximate decoherence factor in
Eq. (\ref{eq:4-9}) are shown in Fig. 4(a) with solid lines and dashed
lines, respectively. For the summation of the series of Bessel functions
in Eq. (\ref{eq:4-9}), we need to set a cutoff of the $\gamma$.
Here, we set a parameter
\begin{equation}
\eta=\frac{4\pi^{2}}{N^{2}\beta\Delta_{e}'}\label{eq:4-17}
\end{equation}
to determine the cutoff as
\begin{equation}
\exp\left(-\eta\gamma_{\mathrm{cutoff}}^{2}\right)=10^{-2}.\label{eq:4-18}
\end{equation}
The parameters are chosen as $N=80,\kappa=10^{-13}N/s,R=0.5\mu m,m=40m_{p}$. The temperatures are respectively $T=483nK, 121nK,31nK$ to guarantee $r=0.5,1,2$ for red, blue and purple lines. And the cutoff $\gamma_\mathrm{cutoff}=9,5,3$ for red, blue and purple dashed lines respectively. The approximate solution describes the decoherence process quite well for the low temperature case, where only few Bessel functions are involved contributing to the oscillating behavior of the decoherence factor.

For the relatively high temperature case such as $r<1$, Eq. (\ref{eq:4-11}) is valid to describe the decoherence process.
The contrast of the exactly decoherence factor obtained
from Eq. (\ref{eq:4-6}) and the approximate decoherence factor in
Eq. (\ref{eq:4-11}) are shown in Fig. 4(b) with solid lines and dashed
lines, respectively. The parameters are as same as ones used for Fig. 4(a). The decoherence processes for short time can
be described quite well by Eq. (\ref{eq:4-11}), while the long time
behavior deviate from the approximate solution because of the linear
dependence of the energy difference we assumed in Eq. (\ref{eq:4-10}).
For $\gamma>1$ the multi period contributions introduce the oscillating
behavior into the decoherence factor.

Besides the temperature, the decoherence time can be elongated by adjusting other parameters such as particle
number $N$, spring constant $\kappa$, the
radius of the ring container $R$ and the particle mass $m$. The numerical calculation directly
based on the exact solution is shown in Fig. 5. The basic parameters
are chosen as $N=80,T=10^{-5}K,\kappa=10^{-13}N/s,R=1\mu m,m=4m_{p}$
with $m_{p}$ is mass of proton. From (a)-(e), the evolutions of the
decoherence factor are depicted for different particle number $N$,
the temperature $T$, spring constant $\kappa$, the radius of the
ring container $R$ and the particle mass $m$. The spontaneous decoherence
occur at first place and it is possible for the decoherence factor
to revive to a relative large quantity at a later time. In some cases
the revival can reach almost 1 as shown in Fig.~\ref{fig:fig5}(b).
Such revival of the decoherence factor results from the contributions
from different periods shown in Fig. ~\ref{fig:fig3}(d), which possibly
cancel each other and eventually elongate the decoherence time. If
we define the decoherence time before the first possible revival,
obviously it is elongated when decreasing the particle number and
the temperature or increasing the spring constant, the ring container
radius and the particle mass. Intriguingly, if the linear mass density
$\eta=N/2\pi R$ is kept unchanged and the particle number increases
just as shown in Fig.~\ref{fig:fig5}(f), the decoherence time is
elongated instead of being shortened when only the particle number
is increased shown in Fig.~\ref{fig:fig5}(a). It implies that the
spontaneous decoherence vanishes at the thermodynamical limit, which
is consistent with the textbook example of phonon in the solid state
physics.
\section{\label{sec:five}CONCLUSION}

We study the spontaneous decoherence of coupled harmonic oscillators
confined in a ring container, where the nearest-neighbor harmonic
potentials are taken into consideration. Without any surrounding environment,
the quantum superposition state prepared in the relative degrees of
freedom gradually loses its quantum decoherence. We study the spontaneous
decoherence existing as the same in the closed multi-particle system
when the symmetry is not broken.

The multi-particle system we study actually possesses s $U(1)\otimes C_{\mathrm{n}}$
symmetry. The Hamiltonian can be divided into the center-of-mass motion
part and the relative motions part. The harmonic potentials between
oscillators are periodic because of the ring configuration. Then nontrivial
boundary conditions emerge to guarantee the single valuedness of the
wave function, which eventually results in that the total energy spectrum
not only depends on the excitations of the relative motion, but also
on the total momentum corresponding to the center-of-mass motion.
The consequence of the nontrivial boundary conditions is adding an
additional phase factor in Eq. (\ref{eq:2-15}), which actually is
equivalent to introducing a gauge field onto the relative motions. There is thin spectrum of the total momentum that
contributes to the decoherence process. If the center-of-mass motion
is not condensed to the state with single momentum, the spontaneous
decoherence process occurs in the superposition states of the relative
motions. Since there is no environment or symmetry breaking field
at all, the decoherence in our model is definitely spontaneous.

This spontaneous decoherence is interpreted by the hidden coupling
between the center-of-mass and relative degrees of freedoms. The paradox
that the information represented by the coherence is always losing
in a closed system can be explained by the infinite degrees of freedom
of the center-of-mass motion acting like a heat bath. Especially,
the spontaneous decoherence completely vanishes at the thermodynamical
limit because the nontrivial boundary conditions become trivial Born-von Karman boundary condition. Our investigation shows that a thermal macroscopic object with
certain symmetries has chance to degrade its quantum properties even
without applying an external symmetry breaking field or surrounding
environment.

\appendix

\section{Solutions of Wave Vectors}

To obtain the wave vectors $\mathbf{q}=(q_{1},q_{2,}\ldots,q_{N-1})^{T}$,
we need to solve the Eq. (\ref{eq:3-6}). Here, $\mathbf{I}=(1,1,\ldots,1)^{T}$
and $L=2\pi R$ is the perimeter of the ring container. The matrix$\mathbf{M}$
in Eq. (\ref{eq:3-6}) is determined by the Fourier transforation
as Eq. (\ref{eq:2-4}). Both the explicity forms of $\mathbf{M}$
for odd and even number $N$ can be unified written as
\begin{equation}
\mathbf{M}=\sqrt{\frac{2}{N}}\left[\begin{array}{cc}
A & B\\
A^{*} & B^{*}
\end{array}\right].\label{eq:Aa-1}
\end{equation}
Takeing the odd number $N$ case as an example, the block matrices
respectively are\begin{subequations}
\begin{eqnarray}
A & =\left[\begin{array}{c}
C_{1}\\
C_{2}\\
\vdots\\
C_{\frac{N-1}{2}}
\end{array}\right] & ,\mbox{ }A^{*}=\left[\begin{array}{c}
C_{\frac{N-1}{2}}\\
C_{\frac{N-3}{2}}\\
\vdots\\
C_{1}
\end{array}\right],\label{eq:Aa-2-1}\\
B & =\left[\begin{array}{c}
S_{1}\\
S_{2}\\
\vdots\\
S_{\frac{N-1}{2}}
\end{array}\right] & ,\mbox{ }B^{*}=-\left[\begin{array}{c}
S_{\frac{N-1}{2}}\\
S_{\frac{N-3}{2}}\\
\vdots\\
S_{1}
\end{array}\right],\label{eq:Aa-2-2}
\end{eqnarray}
\end{subequations}with row vectors\begin{subequations}
\begin{eqnarray}
C_{n} & = & \left[\begin{array}{cccc}
\cos\left(n\phi\right) & \cos\left(2n\phi\right) & \cdots & \cos\left(\left(\frac{N-1}{2}\right)n\phi\right)\end{array}\right],\label{eq:Aa-3-1}\\
S_{n} & = & \left[\begin{array}{cc}
-\sin\left(\left(\frac{1}{2}+n\right)\phi\right) & \sin\left(2\left(\frac{1}{2}+n\right)\phi\right)\end{array}\right.\nonumber \\
 &  & \left.\begin{array}{cc}
\cdots & \left(-1\right)^{\left(\frac{N-1}{2}\right)}\sin\left(\left(\frac{N-1}{2}\right)\left(\frac{1}{2}+n\right)\phi\right)\end{array}\right],\label{eq:Aa-3-2}
\end{eqnarray}
\end{subequations}$(n=1,2,\ldots,\frac{N-1}{2})$ and $\phi=2\pi/N.$
According to identities
\begin{equation}
\sum_{n=1}^{\frac{N-1}{2}}\cos\left(nj\phi\right)=const.\label{eq:Aa-4}
\end{equation}
for any $j=1,2,\ldots,\frac{N-1}{2}$ and the fact that the wave vectors
$\mathbf{q}=(q_{A},q_{B})^{T}$ can be divided into two parts according
to the dimension of the block matrices, the only possible solution
is $q_{A}=\left(q,q,\ldots,q\right)^{T}$ and $q_{B}=\left(0,0,\ldots,0\right)^{T}$.
Therefore, the Eq. (\ref{eq:3-6}) is simplified as
\begin{equation}
qL\sqrt{\frac{2}{N}}\sum_{n=1}^{\frac{N-1}{2}}\cos\left(nm\phi\right)+\frac{2\pi n}{N}=0,\label{eq:Aa-5}
\end{equation}
from which we find the solution $q=\frac{\sqrt{2}n}{\sqrt{N}R}.$

The same procedure can be applied to the case of even number N case
and the solution is a little different from the odd number $N$ case
as $q_{A}=\left(q,q,\ldots,q,q/2\right)^{T}$ and $q_{B}=\left(0,0,\ldots,0\right)^{T}$.

\section{Effective Gauge Fields on Relative Motions}

The total momentum actually plays the role of the effective gauge
field on the relative motions. Starting from the wavefunction obeying
the Floquet theorem as Eq. (\ref{eq:3-1}), the original Schrodiger
equation of the $k$-th relative motion
\begin{eqnarray}
H_{k}\chi_{k}\left(X_{k}\right) & = & \epsilon_{k}\chi_{k}\left(X_{k}\right)\label{eq:Ac-1}
\end{eqnarray}
can be transformed to the Schrodinger equation of the periodic part
as
\begin{equation}
H_{k}^{eff}u_{k}\left(X_{k}\right)=\epsilon_{k}u_{k}\left(X_{k}\right),\label{eq:Ac-2}
\end{equation}
with the exactly same eigenenergy $\epsilon_{k}.$ Here, the effective
Hamiltonian is obtained by a unitary transformation of the original
one as
\begin{align}
H_{k}^{eff} & =e^{-iq_{k}X_{k}}H_{k}e^{iq_{k}X_{k}}\nonumber \\
 & =\frac{\left(P_{k}+\hbar q_{k}\right)^{2}}{2m}+\frac{\kappa}{2}\left(2\sin\frac{\pi k}{N}\right)^{2}X_{k}^{2}.\label{eq:Ac-3}
\end{align}
Apprently the wave vector $q_{k}$ shifts the momentum of the relative
motion, which is equivalent to an $U(1)$ gauge field. Since the wavevector
$q_{k}$ linearly depends on quantum number $n$ as well as the total
momentum $P_{0},$ such gauge fields on relative motions exactly results
from the nonzero total momentum of the system.

\section{Energy Spectrum of The Periodic Harmonic Oscillator }

The Schrodinger equation of the $k$-th relative motions given in
Eq. (\ref{eq:3-10}) is described by a periodic harmonic oscillator
with periodicity
\begin{equation}
\chi_{k}\left(X_{k}+LM_{1}^{k}\right)=e^{iq_{k}LM_{1}^{k}}\chi_{k}\left(X_{k}\right).\label{eq:Ab-1}
\end{equation}
The basic idea to solve the energy spectrum in a periodic potential
is solving the Schrodinger equation in a period and its adjacent period,
then the wavefunctions at the interface of these two periods should
satisfy the continuous condition as Eq. (\ref{eq:3-11}).

The wavefunction of the $k$-th relative mode is the linear combination
of the two degenerate Kummer or confluent hypergeometric functions~\cite{Jr08}
as\begin{subequations}
\begin{align}
f_{\mathrm{e}}(X_{k}) & =\exp\left(-\frac{\xi_{k}^{2}X_{k}^{2}}{2}\right){}_{1}F_{1}\left[\frac{1}{4}(1-\frac{2}{\hbar\omega_{k}}\epsilon_{k});\frac{1}{2};\xi_{k}^{2}X_{k}^{2}\right],\label{eq:Ab-2-1}\\
f_{\mathrm{o}}(X_{k}) & =\xi_{k}r_{k}\exp\left(-\frac{\xi_{k}^{2}X_{k}^{2}}{2}\right){}_{1}F_{1}\left[\frac{1}{4}(3-\frac{2}{\hbar\omega_{k}}\epsilon_{k});\frac{3}{2};\xi_{k}^{2}X_{k}^{2}\right],\label{eq:Ab-2-2}
\end{align}
\end{subequations}with frequencies $\omega_{k}=2\sqrt{\kappa/m}\left|\sin\left(k\pi/N\right)\right|$
and $\xi_{k}=\sqrt{m\omega_{k}/\hbar}$. Here, the subindices $\mathbf{e}$
and $\mathbf{o}$ represent the even and odd parity, respectively.
In contrast to the eigenenergy of the regular harmonic oscillator,
the eigenenergy of the periodic harmonic oscillator $\epsilon_{k}$
is no longer the integer times of the frequencies $\hbar\omega_{k}$.
Consequently, the wavefunction of the $k$-th relative modes within
the coordinate range $X_{k}/LM_{1}^{k}\in\left[-1/2,1/2\right]$ is
assumed to be
\begin{equation}
\chi_{k}(X_{k})=Af_{\mathrm{e}}(X_{k})+Bf_{\mathrm{o}}(X_{k})\label{eq:Ab-3}
\end{equation}
with undetermined coefficients $A$ and $B$. Thus in the next period
$X_{k}/LM_{1}^{k}\in\left[1/2,3/2\right]$, according to Eq.(\ref{eq:Ab-1})
the wavefunction can be written as
\begin{equation}
\chi_{k}(X_{k}+LM_{1}^{k})=e^{iq_{k}LM_{1}^{k}}\left[Af_{\mathrm{e}}(X_{k})+Bf_{\mathrm{o}}(X_{k})\right].\label{eq:Ab-4}
\end{equation}
The continuous conditions require both the wavefunction and derivative
of the wavefunction is continuous as shown in Eq. (\ref{eq:3-11}).
Since the coefficients $A$ and $B$ can not be zero simultaneously,
the determinant of the coefficients matrix of \{$A,B$\} should be
zero as
\begin{equation}
\left|\begin{array}{cc}
f_{\mathrm{e}}(-\frac{l}{2})-e^{i\theta_{k}}f_{\mathrm{e}}(\frac{l}{2}) & f_{\mathrm{o}}(-\frac{l}{2})-e^{i\theta_{k}}f_{\mathrm{o}}(\frac{l}{2})\\
f_{\mathrm{e}}'(-\frac{l}{2})-e^{i\theta_{k}}f_{\mathrm{e}}'(\frac{l}{2}) & f_{\mathrm{o}}'(-\frac{l}{2})-e^{i\theta_{k}}f_{\mathrm{o}}'(\frac{l}{2})
\end{array}\right|=0\label{eq:Ab-5}
\end{equation}
with $l=LM_{1}^{k}$, $\theta_{k}=q_{k}l$ and $f'(a)\equiv\left.\frac{d}{dX}f(X)\right|_{X=a}$.
Finaly we can obtain the constrain for the energy $\epsilon_{k}$
as
\begin{equation}
f_{\mathrm{o}}(\frac{l}{2})f_{\mathrm{e}}'(\frac{l}{2})\cos^{2}\frac{\theta_{k}}{2}+f_{\mathrm{e}}(\frac{l}{2})f_{\mathrm{o}}'(\frac{l}{2})\sin^{2}\frac{\theta_{k}}{2}=0,\label{eq:Ab-6}
\end{equation}
where we have used the parity of the functions $f_{\mathrm{e}}\left(X\right)$
and $f_{\mathrm{o}}\left(X\right)$ to simplify the Eq. (\ref{eq:Ab-5}).
Whether the energy spectrum depends on the total momentum or not relies
on $\theta_{k}\neq0$. Obviously, for those relative motion $k>N/2$
their energy spectrum is independent of the total momentum and thus
have no contribution to the decoherence process.

The energy spectrum $\epsilon_{k}=\left(n_{k}+1/2\right)\hbar\omega_{k}$
depends on both the phase factor $\theta_{k}$ and the dimensionless
parameter $\xi_{k}l$, which is shown in Fig. A1. In Fig. A1(a), the
dimensionless parameter is chosen as $\xi_{k}l=5$ and the particle
number is $N=100.$ Definitely, the energy spectrum depends on the
phase factor $\theta_{k}=q_{k}l.$ For those relative motions with
$q_{k}=0,$ the energy spectrum is only determined by the dimensionless
parameter, which is determined by the geometry of the ring container
and the spring constant. However, for those relative motions with
$q_{k}=\frac{\sqrt{2}n}{\sqrt{N}R},$ the energy spectrum is not only
depends on the total momentum now, but also form a group of thin spectrum
when the total momentum chooses its possible values. In Fig. A1(b),
the phase factor is chosen as $\theta_{k}=\pi/2$ and the particle
number is $N=100.$ By confining the particles in a smaller ring container
via decreasing $\xi_{k}l$, the energy spectrum deviates from the
energy spectrum of standard harmonic oscillator greatly. When $\xi_{k}l\gg1$
the energy spectrum is almost coincide with the standard one, which
means the affect of the phase factor is also suppressed for a larger
ring container or weak spring constant.

\begin{figure}[ptb]
\begin{centering}
\includegraphics[bb=41 537 571 775,clip,width=3.5in]{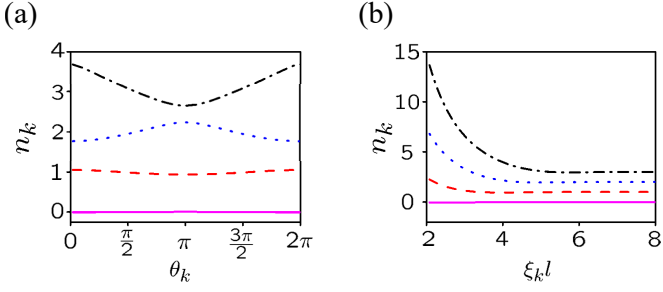}
\par\end{centering}

\caption{(Color online) (a) Energy spectrum $n_{k}$ versus phase factor $\theta_{k}$.
The parameters are chosen as $N=100,\xi_{k}l=5$. (b) Energy spectrum
$n_{k}$ versus the periodicity $\xi_{k}l$. The parameters are chosen
as $N=100,\theta_{k}=\pi/2$. The black dotdashed line, blue dotted
line, red dashed line and the magenta solid line represent the first
four eigenstates of the periodic harmonic oscillator. Definitely,
the energy spectrum varies with the $\theta_{k}$, implying the dependence
of the total momentum. Moreover, the periodic harmonic oscillator
becomes normal one when $\xi_{k}l\gg1$ whatever $\theta_{k}$ is.}

\label{fig:figA1}
\end{figure}


We rewrite the Eq.(\ref{eq:Ab-6}) as
\begin{equation}
\tan^{2}\frac{\theta_{k}}{2}=-F\left(l,\epsilon_{k}\right)\label{eq:Ab-7}
\end{equation}
with
\begin{equation}
F\left(l,n_{k}\right)=\frac{f_{\mathrm{o}}(\frac{l}{2})f_{\mathrm{e}}'(\frac{l}{2})}{f_{\mathrm{e}}(\frac{l}{2})f_{\mathrm{o}}'(\frac{l}{2})}.\label{eq:Ab-8}
\end{equation}
To obtain the approximate energy spectrum which depends linearly on
the total momentum, we expand the Eq.(\ref{eq:Ab-7}) at the vicinity
of the phase factor $\theta_{k}=\left(\frac{1}{2}+\mu\right)\pi$
and $\xi_{k}l\apprge1.$ In this sense, the approximate energy spectrum
is obtained as
\begin{equation}
\epsilon_{k}\left(n,\alpha\right)=\left(\frac{1}{2}+\alpha'+\delta_{k}\left(n,\alpha\right)\right)\hbar\omega_{k},\label{eq:Ab-9}
\end{equation}
where $\alpha'$ is the solution of $F\left(l,\alpha'\right)=-1$,
$\alpha=0,1,\ldots$ is non-negative integer number and deviation
\begin{equation}
\delta_{k}\left(n,\alpha\right)=g_{0}\left(\alpha\right)+g_{1}\left(k,\alpha\right)n\label{eq:Ab-10}
\end{equation}
with coefficients\begin{subequations}
\begin{align}
g_{0}\left(\alpha\right) & =-\frac{1+F\left(l,\alpha\right)+\left(-1\right)^{\mu}\left(1+2\mu\right)\pi}{G\left(l,\alpha\right)},\label{eq:Ab-11-1}\\
g_{1}\left(k,\alpha\right) & =\left(-1\right)^{\mu}2\pi\frac{M_{1}^{k}}{G\left(l,\alpha\right)}\frac{\sqrt{2}}{\sqrt{N}},\label{eq:Ab-11-2}
\end{align}
\end{subequations}and function $G\left(l,\alpha\right)\equiv\left.\frac{d}{dn_{k}}F\left(l,n_{k}\right)\right|_{n_{k}=\alpha'}$
is the derivative of the function $F\left(l,n_{k}\right).$

\begin{acknowledgements}

The author thank H. C. Fu for helpful discussion. This work is supported by NSFC Grants No. 11504241 and the Natural
Science Foundation of SZU Grants No. 201551.

\end{acknowledgements}

\end{document}